\documentclass{article_saj}
\usepackage{graphicx,saj,multicol,subeqnarray}
\usepackage{natbib}
\usepackage{float}
\usepackage{xcolor}
\usepackage{widetext}
\usepackage{url}
\usepackage{bm}
\usepackage{tikz} 
\usepackage{pifont} 
\usepackage{amsfonts}
\usepackage{amssymb}
\usepackage{amsmath,upgreek}
\usepackage{titlesec}
\titlelabel{\thetitle.\quad}
\definecolor{xlinkcolor}{cmyk}{1,0.6,0,0}
\usepackage[bookmarks=false,         
     pdfnewwindow=true,      
     colorlinks=true,    
     linkcolor=xlinkcolor,     
     citecolor=xlinkcolor,     
     filecolor=xlinkcolor,  
     urlcolor=xlinkcolor,      
final=true
]{hyperref}
\usepackage{changepage}
\usepackage{diagbox}
\usepackage{stackengine}
\newcommand\xrowht[2][0]{\addstackgap[.5\dimexpr#2\relax]{\vphantom{#1}}}

\def\udc{52}

\begin{document}
\parindent=.5cm
\baselineskip=3.8truemm
\columnsep=.5truecm
\newenvironment{lefteqnarray}{\arraycolsep=0pt\begin{eqnarray}}
{\end{eqnarray}\protect\aftergroup\ignorespaces}
\newenvironment{lefteqnarray*}{\arraycolsep=0pt\begin{eqnarray*}}
{\end{eqnarray*}\protect\aftergroup\ignorespaces}
\newenvironment{leftsubeqnarray}{\arraycolsep=0pt\begin{subeqnarray}}
{\end{subeqnarray}\protect\aftergroup\ignorespaces}
%


\markboth{\eightrm Light Curve Solution} 
{\eightrm S. WADHWA {\lowercase{\eightit{et al.}}}}

\begin{strip}

{\ }

\vskip-1cm

\publ


{\ }


\title{Effective Temperature and the Light Curve Solution of Contact Binary Systems }


\authors{S. Wadhwa$^{1}$, N.F.H Tothill$^{1}$, M.D Filipovi{\' c}$^{1}$ and  A.Y De Horta$^{1}$}

\vskip3mm


\address{$^1$School of Science, Western Sydney University,\break Locked Bag 1797, Penrith, NSW 2751, Australia.}


\Email{19899347@student.westernsydney.edu.au}







\summary{With increasing number of contact binary discoveries and the recognition that luminous red novae are the result of contact binary merger events, there has been a significant increase in the number of light curve solutions appearing in the literature. One of the key elements of such solutions is the assignment and fixing of the effective temperature of the primary component ($\rm T_1$). Sometimes much is made of the assigned value with expectation of significantly different light curve solutions even though theoretical considerations suggest that absolute value of $\rm T_1$ has little influence on the geometric elements of the light curve solution. In this study we show that assigning $\rm T_1$ over a range of $1000K$ has no significant influence on the light curve solutions of two extreme low mass ratio contact binary systems. In addition, we explore the use of photometric spectral energy distribution as a potential standard for assigning $\rm T_1$ in the absence of spectroscopic observations.}



\end{strip}

\tenrm


\section{Introduction}

\indent

\footnotetext[0]{\copyright \ 2023 The Author(s). Published by
Astronomical Observatory of Belgrade and Faculty of Mathematics,
University of Belgrade. This open access article is distributed under
CC BY-NC-ND 4.0 International licence.}

Numerous all sky surveys have resulted in a massive proliferation in the identification of contact binary systems with approximately half a million catalogued in the International Variable Star Index (VSX) \citep{2006SASS...25...47W}. With such proliferation there has been a corresponding increase in published analysis of contact binary light curves. It is known that the shape of contact binary light curves, particularly those exhibiting total eclipses, are almost entirely dependent on three main geometric parameters namely the inclination ($i$), the mass ratio ($q$) and degree of contact or fill-out ($f$) \citep{1993PASP..105.1433R, 2001AJ....122.1007R, 2022JApA...43...94W}. There is strong correlation between these parameters and successful light curve analysis in the absence of radial velocity measurements can only be achieved if some constraints can be placed due to the shape of the light curve itself. The presence of total eclipses provides the morphological features that place a strong constrain on the [$q,i$] domain. In the presence of total eclipses one can systematically search the [$q,i$] domain to find a set of geometric parameters that yield the best fit between observed and modelled light curves \citep{2005Ap&SS.296..221T}.

Even though the shape of the light curve, especially when total eclipses are not present, may be influenced by the absolute value of the component temperatures \citep{2020Galax...8...57W}, the presence of complete eclipses and the associated constraints on the geometric parameters greatly overshadows these variations such that temperature variations of a few hundred degrees are thought not to influence the light curve modelled geometric parameters. During analysis of contact binary light curves the temperature of the secondary ($T_2$), $f$, $i$ and the dimensionless luminosity of the primary ($L_1$) are regarded as the adjustable parameters while the temperature of the primary component ($\rm T_1$) is fixed. There is no standard method for the assigning of $\rm T_1$ and wide variations in the effective temperature can be found depending on the colour or spectral classification used. In a recent study reporting photometric analysis of twelve extreme low mass ratio contact binaries, the authors reported wide variations in effective temperature of the primary based on various catalogued colour and spectral calibrations \citep{2023PASP..135g4202W}. 

Given the wide variations in effective temperature, particularly with colour based estimates, many investigators are increasingly using low resolution spectral classification as a guide to assigning $\rm T_1$. Unfortunately, even for bright examples, spectral observations still require access to mid-level equipment which may not be readily available. As noted above, numerous all sky surveys have been undertaken recently covering a wide range of the electromagnetic spectrum from ultraviolet to infrared. As described by \citep{2007ApJS..169..328R, 2008A&A...492..277B} it is possible to collectively examine the multi-band photometry as a single Spectral Energy Distribution (SED) which can then be fitted to modelled synthetic spectra to estimate the effective temperature. Using isolated examples few investigators have shown good correlation between spectral class and SED determined effective temperature \citep{2022ApJ...927...12P, 2023arXiv230809998W}. 

Although theoretical framework \citet{2009ebs..book.....K} suggests that the geometric light curve solution will not differ with absolute temperature values we can find no formal published study that demonstrates this practically. Also, as noted above, only isolated examples exist demonstrating the utility of SED in assigning effective temperature. In this study we explore both of these issues. Firstly, we compare the spectral class determined temperature of the primary with the SED determined temperature for contact binaries from spectral class F3 to K4. Secondly, we undertake detailed modelling, of the V and R band light curves, of two totally eclipsing contact binary with temperature of the primary fixed using either spectral class or \ SED $\pm$ 100, 300 and 500K to confirm that geometric solutions do not change significantly with changing absolute temperature values.

\section{Spectral Class and SED}

We selected 12 bright contact binaries with determined spectral class from The Large Sky Area Multi-Object Fiber Spectroscopic Telescope (LAMOST) survey \citep{2018yCat.5153....0L}. We chose relatively bright examples as these were more likely to have been included in many of the photometric surveys. We determined the spectral class based effective temperature using the April 2022 update of \citet{2013ApJS..208....9P} calibration tables of spectral class and temperature for main sequence stars. We constructed collective photometric SEDs for each star as described by \citet{2008A&A...492..277B}. All SEDs were then fitted to modelled synthetic spectra which incorporated Kurucz atmospheres using ${\chi}^2$ minimisation as a goodness of fit parameter. Comparison of spectral class and SED determined effective temperature is summarised in Table 1 and representative SED and fitted modelled spectra are illustrated in Figure 1. As can be seen from Table 1 there is good agreement between SED and spectral class determined effective temperature and we consider SED as a more robust alternative to single colour calibration for assigning $\rm T_1$ when spectral observations are lacking.

  \begin{table*}

   \centering
   \begin{tabular}{|l|l|l|l|}
    \hline\xrowht[()]{10pt}
        \hfil Star&\hfil Sp Class  &\hfil Sp $T_{eff}$ (K) &\hfil SED $T_{eff}$ (K)\\\hline\xrowht{10pt}
        \hfil CRTS J090136.9+443723&\hfil F3&\hfil 6750 &\hfil 6500\\\hline\xrowht{10pt}
        \hfil PQ Leo&\hfil F4&\hfil6670&\hfil 6750\\\hline\xrowht{10pt}
        \hfil NU Boo&\hfil F5&\hfil 6550&\hfil 6250\\\hline\xrowht{10pt} 
        \hfil V356 Dor&\hfil F6&\hfil 6350&\hfil 6250\\\hline\xrowht{10pt}
        \hfil NN Lyn &\hfil F7&\hfil 6280&\hfil 6000\\\hline\xrowht{10pt}
        \hfil EI CMi&\hfil F8&\hfil6180&\hfil6000\\\hline\xrowht{10pt}
        \hfil OR Leo&\hfil G0&\hfil5930&\hfil6000\\\hline\xrowht{10pt}
        \hfil KR Lyn&\hfil G2&\hfil5770&\hfil5750\\\hline\xrowht{10pt}
        \hfil GSC 02992-01147&\hfil G4&\hfil5680&\hfil5500\\\hline\xrowht{10pt}
        \hfil OV Leo&\hfil G7&\hfil5550&\hfil5250\\\hline\xrowht{10pt}
        \hfil HZ CVn&\hfil G9&\hfil5380&\hfil5250\\\hline\xrowht{10pt}
        \hfil V625 And&\hfil K1&\hfil5170&\hfil5000\\\hline\xrowht{10pt}
        \hfil ASAS J020753+2034.1&\hfil K4&\hfil4600&\hfil4750\\\hline

    \end{tabular}
    \caption{Comparison of effective temperatures of contact binaries based on spectral class and spectral energy distribution from F3 to K4.} 
   
    \end{table*}

     \begin{figure*}[!ht]
    \label{fig:F1}
	\includegraphics [width=\textwidth] {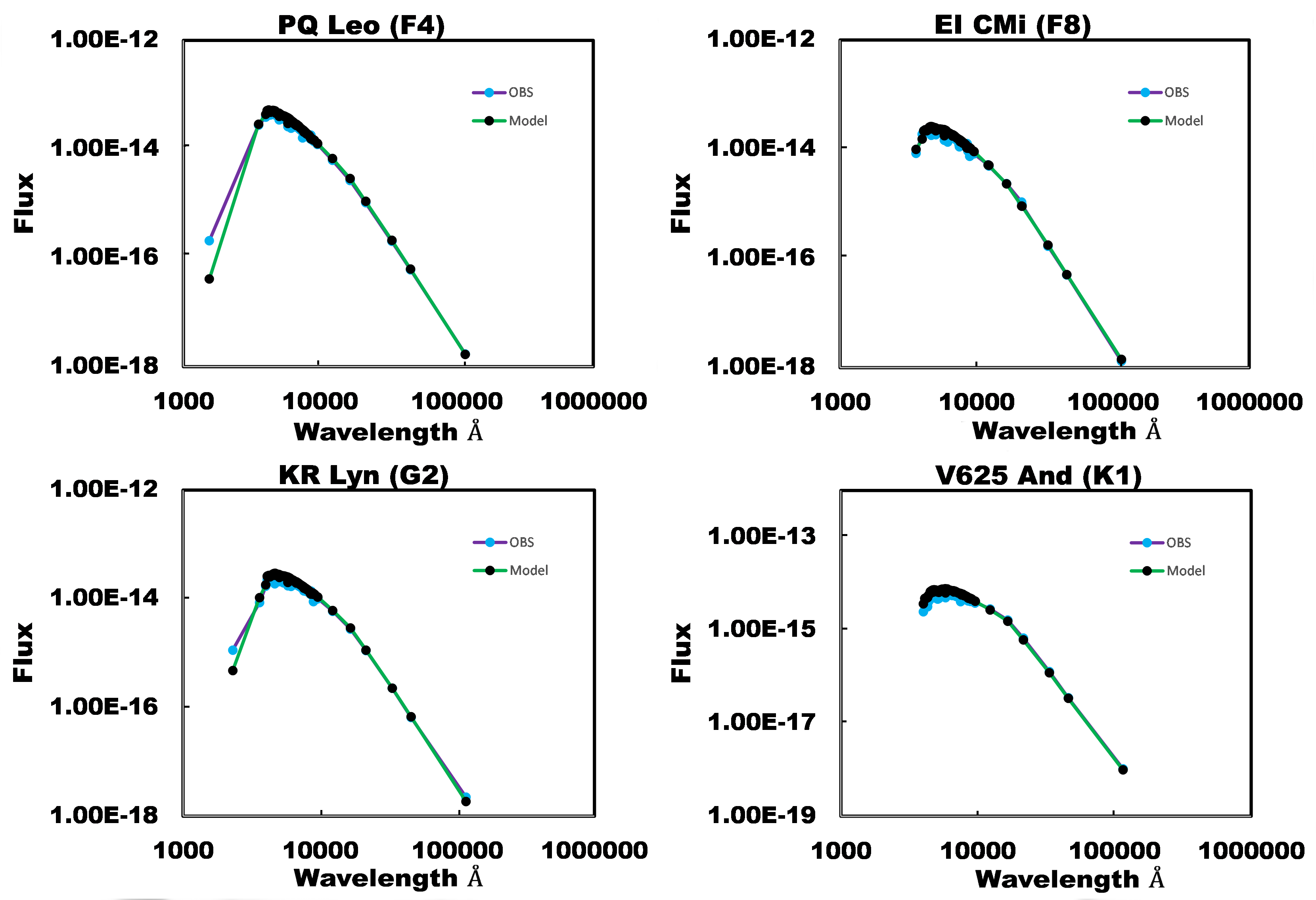}
    \caption{Observed and fitted SEDs for contact binaries ranging from spectral class F4 to K1. The observed photometry is indicated in purple and the fitted model in green. The flux on the vertical axes is in erg/cm$^2$/s/\AA. The wavelength is in Angstroms (\AA). Both axes are in log scale.}
    \end{figure*}

      \begin{figure*}
    \label{fig:F2}
	\includegraphics [width=\textwidth] {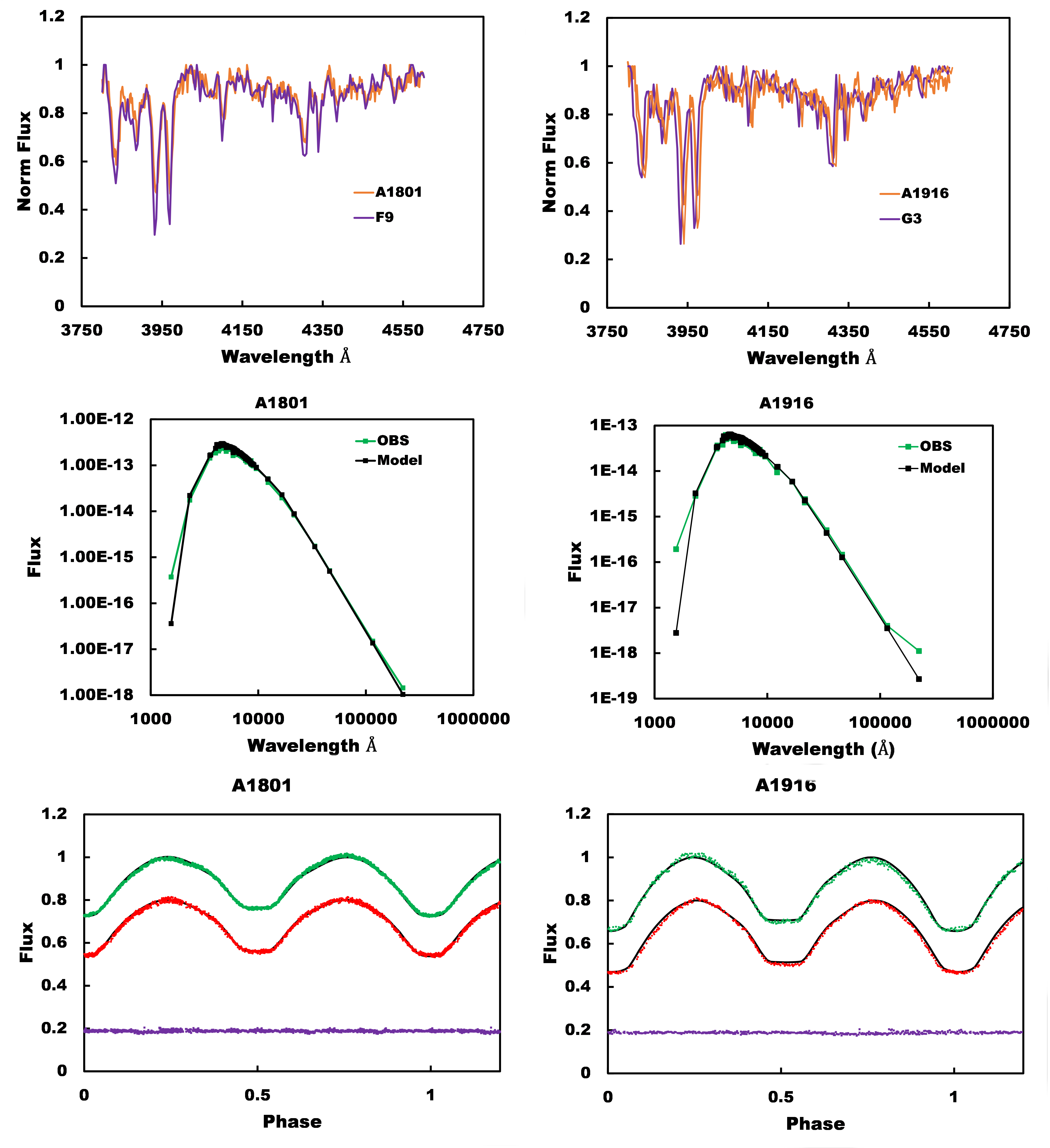}
    \caption{Spectra, SEDs and fitted light curves for A1801 and A1906. SED flux on the vertical axes is in erg/cm$^2$/s/\AA. The wavelength is in Angstroms (\AA). Both axes are in log scale.}
    \end{figure*}

\begin{table*}[!ht]

   \centering
   \begin{tabular}{|l|l|l|l|l|l|l|l|}
    \hline
        \hfil\diagbox{LC}{\\$\rm T_1$}  &\hfil 5750K &\hfil 5950K &\hfil 6150K &\hfil \textbf{6250K}&\hfil 6350K&\hfil 6550K&\hfil 6750K\\\hline\xrowht[()]{10pt}
        \hfil $T_2$(K)&\hfil $5554\pm10$&\hfil $5729\pm10$&\hfil$5933\pm10$&\hfil$\mathbf{6024\pm10}$&\hfil $6127\pm12$&\hfil$6310\pm11$&\hfil$6510\pm10$\\\hline\xrowht{10pt}
        \hfil$T_2/T_1$&\hfil0.97&\hfil0.96&\hfil0.96&\hfil \textbf{0.96}&\hfil0.96&\hfil0.96&\hfil0.96\\\hline\xrowht{10pt}
        \hfil Incl.($^\circ$)&\hfil$76.5\pm0.3$&\hfil$76.2\pm0.3$&\hfil$76.0\pm0.3$&\hfil$\mathbf{77.0\pm0.5}$&\hfil$77.2\pm0.4$&\hfil$77.3\pm0.4$&\hfil$77.4\pm0.5$\\\hline\xrowht{10pt} 
        \hfil Fillout (\%)&\hfil$34\pm5$&\hfil$34\pm5$&\hfil$35\pm4$&\hfil$\mathbf{34\pm4}$&\hfil$35\pm5$&\hfil$34\pm4$&\hfil$33\pm4$\\\hline\xrowht{10pt}
        \hfil Mass Ratio ($q$) &\hfil$0.151$&\hfil$0.153$&\hfil$0.151$&\hfil$\mathbf{0.151}$&\hfil$0.150$&\hfil$0.151$&\hfil$0.151$\\\hline\xrowht{10pt}
        \hfil$r_1$ (mean)&\hfil0.559&\hfil0.558&\hfil0.559&\hfil\textbf{0.559}&\hfil0.560&\hfil0.559&\hfil0.559\\\hline\xrowht{10pt}
        \hfil$r_2$(mean)&\hfil0.245&\hfil0.246&\hfil0.245&\hfil\textbf{0.245}&\hfil0.245&\hfil0.245&\hfil0.244\\\hline

    \end{tabular}
    \caption{A1801 Light curve (LC) solution for different values of $\rm T_1$ from 5750K to 6750K. $T_2/T_1$ represents the component temperature ratio and $r_{1,2}$ are the geometric mean of the fractional radii. The error for the mass ratio was $\pm0.002$ in all cases. The central highlighted solution corresponds to SED modelled value of $\rm T_1$.} 
   
    \end{table*}

    \begin{table*}

   \centering
   \begin{tabular}{|l|l|l|l|l|l|l|l|}
    \hline
        \hfil\diagbox{LC}{\\$\rm T_1$}  &\hfil 5220K &\hfil 5420K &\hfil 5620K &\hfil \textbf{5720K}&\hfil 5820K&\hfil 6020K&\hfil 6220K\\\hline\xrowht[()]{10pt}
        \hfil $T_2$(K)&\hfil $5105\pm20$&\hfil $5293\pm18$&\hfil$5481\pm15$&\hfil$\mathbf{5578\pm15}$&\hfil $5681\pm16$&\hfil$5867\pm16$&\hfil$062\pm20$\\\hline\xrowht{10pt}
        \hfil$T_2/T_1$&\hfil0.98&\hfil0.98&\hfil0.98&\hfil \textbf{0.98}&\hfil0.98&\hfil0.97&\hfil0.97\\\hline\xrowht{10pt}
        \hfil Incl.($^\circ$)&\hfil$89.90^{+0.1}_{-2.2}$&\hfil$90.00^{+0.0}_{-1.8}$&\hfil$89.9^{+0.1}_{-1.4}$&\hfil$\mathbf{89.8^{+0.2}_{-1.5}}$&\hfil$89.8^{+0.2}_{-1.5}$&\hfil$89.9^{+0.2}_{-1.3}$&\hfil$89.7^{+0.3}_{-2.0}$\\\hline\xrowht{10pt} 
        \hfil Fillout (\%)&\hfil$46\pm7$&\hfil$45\pm8$&\hfil$50\pm6$&\hfil$\mathbf{50\pm5}$&\hfil$49\pm5$&\hfil$52\pm5$&\hfil$53\pm5$\\\hline\xrowht{10pt}
        \hfil Mass Ratio ($q$) &\hfil$0.188$&\hfil$0.191$&\hfil$0.190$&\hfil$\mathbf{0.191}$&\hfil$0.191$&\hfil$0.192$&\hfil$0.191$\\\hline\xrowht{10pt}
        \hfil$r_1$ (mean)&\hfil0.547&\hfil0.545&\hfil0.548&\hfil\bf{0.548}&\hfil0.547&\hfil0.548&\hfil0.549\\\hline\xrowht{10pt}
        \hfil$r_2$(mean)&\hfil0.267&\hfil0.2680&\hfil0.271&\hfil\bf{0.271}&\hfil0.271&\hfil0.273&\hfil0.273\\\hline

    \end{tabular}
    \caption{A1916 Light curve (LC) solution for different values of $\rm T_1$ from 5220K to 6220K. $T_2/T_1$ represents the component temperature ratio and $r_{1,2}$ are the geometric mean of the fractional radii. Due to increased scatter the fit for A1916 was not as clean as for A1801 and the error for the mass ratio was $\pm0.004$ in all cases. The central highlighted solution corresponds to spectra modelled value of $\rm T_1$.} 
   
    \end{table*}

 \section{Light Curve Solution and $\rm T_1$}

ASAS J180157-7228.1 (A1801) ($\alpha_{2000.0} = 18\ 01\ 56.66$, $\delta_{2000.0} = -72\ 28\ 07.0$) was recognised a contact binary by the All Aky Automated Survey (ASAS) \citep{2002AcA....52..397P} with a period of 0.355909 days and an amplitude of 0.35 magnitude. The system was observed over 4 night in August 2023 with the Western Sydney University 0.6m telescope equipped with standard Johnson V and R filters and a cooled SBIG 8300T CCD camera. Images were obtained using both V and R filters. All images were calibrated using multiple dark, flat and bias frames. Differential photometry was performed using the AstroImageJ \citep{2017AJ....153...77C} package. TYC 9298-140-1 was used as the comparison star and 2MASS 18023134-7230047 as the check star. The V band amplitude of the system was estimated to be 0.35 mag with brightest magnitude of 10.36 and mid eclipse magnitude of 10.66. Using all available observed and survey V band data we refine the orbital elements as follows:

\begin{center}
    
    $HJD_{min} = 2460150.9825208(406) + 0.3559121(30)E$\\                   

\end{center}

ASAS J191621-6802.3 (A1916) ($\alpha_{2000.0} = 19\ 16\ 21.01$, $\delta_{2000.0} = -68\ 02\ 19.2$) another ASAS discovery with an amplitude of 0.46  magnitude and period 0.364588 days. The system was observed over 5 nights between July and August 2023 with the Western Sydney University 0.6m telescope. Again V and R band images were acquired, calibrated and photometry performed using the AstroImageJ package. TYC 110-123-59 was the comparison star and 2MASS 19155297-6758004 was the check star.  The V band amplitude of the system was estimated to be 0.46 mag (12.03 - 12.49) and mid eclipse magnitude of 12.43. Using all available observed and survey V band data we refine the orbital elements as follows: 

\begin{center}
    
    $HJD_{min} = 2460155.1296218(753) + 0.36459189(25)E$\\                   

\end{center}

Low resolution spectra for each system was obtained using the 2 meter telescopes of the Las Cumbres Observatory  (LCO) network over two nights in August 2023. The LCO is a fully automated network and provides fully calibrated spectra without user input. We compared the LCO spectra against standard library spectra \citep{1984ApJS...56..257J, 1998PASP..110..863P} to assign the spectral class as F9 for A1801 and G3 for A1916. The corresponding temperatures (6050K and 5720K) were interpolated from the April 2022 updated tables from \citep{2013ApJS..208....9P}. We constructed and fitted SEDs for each system as described above. The SED determined temperatures for A1801 was 6250K and 6000K for A1916. The observed and modelled spectra and SEDs are illustrated in Figure 2.

Both light curves demonstrate complete eclipses and hence are suitable for light curve analysis. The light curves were analysed using the 2013 version of the Wilson-Devenney code \citep{1998ApJ...508..308K, 1990ApJ...356..613W}. To fully illustrate the flexibility of assigning the temperature of the primary we fixed the temperature of the primary as 6250K for A1801 (SED based) and 5720K for A1916 (spectral based). We utilised the mass ratio search grid method for fixed value of $q$ from 0.05 to 1.0 in increments of 0.01 to $q=0.1$ and then in increments of 0.02 to $q=1.0$. The adjustable parameters were $T_2$, $L_1$, $f$ and $i$. As the estimated temperature is less than 7200K the gravity darkening coefficients were fixed as ($g_{1,2} = 0.32$), bolometric albedoes were fixed as ($A_{1,2} = 0.5$) and simple reflection treatment was applied. Limb darkening coefficients were interpolated from \citep{1993AJ....106.2096V}. Iterations were carried out until the suggested corrections was less than the reported standard deviation for each adjustable parameter. Once the approximate solution was obtained the $q$ search was further narrowed to 0.001 increments to find the best fit solution.

The above analysis was repeated for various fixed values of $\rm T_1$ at $T_1, T_1\pm100\rm K, T_1\pm300\rm K$, and $T_1\pm500\rm K$ for both systems. In total each system was modelled for 7 different values of $\rm T_1$ in the range $T_1\pm500\rm K$. The geometric elements for each solution are summarised in Tables 2 and 3 and the observed and fitted light curves both systems are illustrated in Figure 2.

    \section{Absolute Parameters}
    
    In the absence of high resolution radial velocity measurements one is reliant of various astrophysical correlations to estimate the absolute parameters of contact binaries. The black body approximation $L = 4\sigma R^2T^4$ is often used to estimate the radius of the primary ($R_1$) from the assigned value of $T_1$ and the observed luminosity (absolute magnitude) of the system. The mass-radius relation for main sequence stars is then used to estimate the mass of the primary and from the light curve solution the mass of the secondary and then Kepler's third law to estimate the separation of the components. Such an approach, although valid for single stars, likely only represents a rough estimate in the case of contact binary systems as it is highly dependent on the assigned temperature. As noted above there can be wide variation in the estimated temperature of the primary and as noted by \citet{2023arXiv230811906W}, a 200K variation in the assigned value of $T_1$ can lead to a greater than 10\% change in the estimated value of $M_1$ for low mass stars. Additionally, a number of steps are required to determine luminosity, radius and then mass, each associated with its own error which would require propagation leading to a larger overall error in the estimate. Lastly, the black body and main sequence approximations are based on a spherical configuration, it is well known that binary star components are extended, distorted and fill their Roche lobes such that the mean radius of both the primary and secondary are considerably larger than their main sequence counterparts \citep{2022JApA...43...94W}.

    \begin{table*}
    \centering
    
    \begin{tabular}{|l|l|l|l|l|l|l|l|}
    \hline
       \hfil  & \hfil $M_{V1}$ &\hfil $M_1/M_{\odot}$ &\hfil $M_2/M_{\odot}$ &\hfil $R_1/R_{\odot}$ &\hfil $R_2/R_{\odot}$ &\hfil $A/R_{\odot}$  \\ \hline
        \hfil A1801 & $4.05\pm0.08$ & $1.11\pm0.02$ & $0.17\pm0.01$ & $1.28\pm0.02$ & $0.56\pm0.01$ & $2.29\pm0.02$  \\ \hline
        \hfil A1916 & $4.39\pm0.06$ & $1.04\pm0.02$ & $0.20\pm0.01$ & $1.26\pm0.02$ & $0.63\pm0.02$ & $2.30\pm0.02$ \\ \hline
    \end{tabular}
    \caption{Absolute parameters for A1801 and A1916.}
\end{table*}

    We favour an observational approach for estimating the mass of the primary. The Two Micron All Sky Survey (2MASS) acquired high precision simultaneous photometry in multiple infrared bands \citep{2006AJ....131.1163S} and the GAIA - EDR3 provides high precision distance estimates, particularly for nearby systems \citep{2022A&A...658A..91A, 2023A&A...674A...1G}. We have previously described \citet{2023PASP..135g4202W} methodology to estimate the mass of the primary as a mean of the 2MASS $J-H$ colour - mass of low mass main sequence stars calibration and the absolute magnitude - main sequence calibration. The apparent magnitude of the secondary eclipse, being total, represents the apparent magnitude of the primary and can be used to estimate its absolute magnitude. As extinction maps estimate extinction to infinity we distance scale the extinction based on GAIA distance to determine absolute magnitude for the primary as described in \citep{2023PASP..135g4202W}. The mass of the secondary can be determined from the mass ratio and separation ($A$) using Kepler's third law. As the systems are in a contact configuration the radii of the components can be estimated as $R_{1,2} = A(r_{1,2})$ where $r_{1,2}$ are the geometric means of the fractional radii from three different orientations provided by the light curve solution. The absolute parameters determined using this methodology are summarised in Table 4. Our preference for the methodology are obvious. We estimate absolute parameters adopting observations and geometric elements of the light curve solution. As we have shown in this report the estimation of absolute parameters are essentially independent of the assigned value of $T_1$ unlike the black body approximations often used. In presence of total eclipses the Roche geometry places tight constrains on the ($q,i$) and ($q,f$) domains such that the light curve solution is essentially the same regardless of the assigned temperature, at least within 500K as demonstrated here.

    \section{Discussion and Conclusion}
Since the confirmation that luminous red novae are the result of merger of contact binary components \citep{2011A&A...528A.114T} combined with ever increasing number of catalogued contact binary systems there has been a proliferation contact binary curve solutions appearing in the literature. The hallmark of achieving a successful light curve solution is the presence of a complete eclipse. During such analysis various, mainly geometric, parameters are adjusted to achieve a good fit between observed and modelled light curves. Absolute value of the component temperatures, at least theoretically, are not thought to have a significant influence on the determination of geometric parameters such as the mass ratio, fractional radii, fill-out and inclination. Theory suggests that where completely eclipsing light curves are present there is tight constrain on the $T_2/T_1$ ratio but not the absolute temperatures \citep{1993PASP..105.1433R, 2001AJ....122.1007R}. This is the first study, that we are aware of, that tests theory through a practical example. We performed detailed modelling of two low amplitude contact binary system with total eclipses. We carried out modelling at various temperatures above a below an estimate of the temperature of the primary. As can be seen from Tables 2 and 3 there is very little difference in the modelled geometric parameters regardless of the fixed temperature assigned to the primary component. As expected there is tight constrain on the $T_2/T_1$ ratio which remained essentially fixed regardless of the assigned value of $\rm T_1$. 

Unless the estimated value of $\rm T_1$ is to be used in further analysis such as luminosity and radius estimations, although these are probably better estimated using distance, absolute magnitude and colour calibrations, a reasonable estimation should suffice to get accurate estimation of geometric parameters which are more important in determining potential orbital stability. Unfortunately even a reasonable estimation of $\rm T_1$ can prove problematic given the very wide variation between different colour and spectral calibrations. Spectra based estimations probably represent a good standard method however obtaining, even low resolution spectra, requires access to modest equipment levels. A possible solution is to use a collective photometric SED which incorporates most available photometric data and model the SED to theoretical spectra. We show in this study that such an approach for low mass (spectral class F to K) is sufficiently accurate and results compare favourably to spectral class estimation of the effective temperature.

In conclusion, we feel that rigorous estimation for the effective temperature of the primary component which is normally fixed during light curve analysis is not required in the case of totally eclipsing systems and where only geometric parameters determined through light curve analysis are required for further analysis. The classical case is the determination of orbital stability. Using the detailed observational methodology described in \citet{2023PASP..135g4202W} we estimate the mass of the primary of A1801 as $1.11\pm0.02M_{\odot}$ and of A1901 as $1.04\pm0.02M_{\odot}$. Using the simplified quadratic relations from \citet{2021MNRAS.501..229W} the orbital instability mass ratio range for A1801 is 0.085 - 0.097 and for A1901 0.095 - 0.110. The modelled mass ratios for both systems are higher than the upper limit of the instability mass ratio and as such both systems would be considered stable. This conclusion does not change regardless of the assigned temperature of the primary as the modelled geometric parameters are not influenced by absolute value of the assigned temperature.


\acknowledgements{Based on data acquired on the Western Sydney University, Penrith Observatory Telescope. We acknowledge the traditional custodians of the land on which the Observatory stands, the Dharug people, and pay our respects to elders past and present.}

{This publication makes use of VOSA, developed under the Spanish Virtual Observatory (https://svo.cab.inta-csic.es) project funded by MCIN/AEI/10.13039/501100011033/ through grant PID2020-112949GB-I00. VOSA has been partially updated by using funding from the European Union's Horizon 2020 Research and Innovation Programme, under Grant Agreement number 776403.}



\newcommand\eprint{in press }

\bibsep=0pt

\bibliographystyle{aa_url_saj}

{\small

\bibliography{sample_saj}
}

\begin{strip}

\end{strip}

\clearpage

{\ }

\clearpage

{\ }

\newpage

\begin{strip}

{\ }






\vskip3mm





















\end{strip}


\end{document}